\documentclass[reqno]{amsart}

\newcommand{\be}{\begin{equation}}
\newcommand{\ee}{\end{equation}}

\newcommand{\pp}{\partial}
\newcommand{\vv}[1]{\boldsymbol{\mathrm{#1}}}

\usepackage{amsmath}
\usepackage{amsthm,amssymb}
\usepackage{bm}
\usepackage{graphicx}

\theoremstyle{remark}

\title[]{A hybrid inversion scheme combining Markov chain Monte Carlo and iterative methods for determining optical properties of random media}

\author[]{Yu Jiang}
\address{School of Mathematics, Shanghai University of Finance and Economics, 
Shanghai 200433, P.R. China}
\email{jiang.yu@mail.shufe.edu.cn}

\author[]{Yoko Hoshi}
\address{Institute for Medical Photonics Research,
Hamamatsu University School of Medicine,
Hamamatsu 431-3192, Japan}
\email{yhoshi@hama-med.ac.jp}

\author[]{Manabu Machida}
\address{Institute for Medical Photonics Research,
Hamamatsu University School of Medicine,
Hamamatsu 431-3192, Japan}
\email{machida@hama-med.ac.jp}

\author[]{Gen Nakamura}
\address{Department of Mathematics, Hokkaido University, 
Sapporo 060-0810, Japan}
\email{nakamuragenn@gmail.com}

\date{\today}

\begin{document}

\begin{abstract}
Near-infrared spectroscopy (NIRS) including diffuse optical tomography is an imaging modality which makes use of diffuse light propagation in random media. When optical properties of a random medium is investigated from boundary measurements of reflected or transmitted light, iterative inversion schemes such as the Levenberg-Marquardt algorithm are known to fail when initial guesses are not close to the true value of the coefficient to be reconstructed. In this paper, we investigate how this weakness of iterative schemes is overcome by the use of Markov chain Monte Carlo. Using time-resolved measurements performed against a polyurethane-based phantom, we present a case that the Levenberg-Marquardt algorithm fails to work but the proposed hybrid method works well. Then with a toy model of diffuse optical tomography we illustrate that the Levenberg-Marquardt method fails when it is trapped by a local minimum but the hybrid  method can escape from local minima by using the Metropolis-Hastings Markov chain Monte Carlo algorithm until it reaches the valley of the global minimum. The proposed hybrid scheme can be applied to different inverse problems in NIRS which are solved iteratively. We find that for both numerical and phantom experiments optical properties such as the absorption and reduced scattering coefficients can be retrieved without being trapped by a local minimum when Monte Carlo simulation is run only about $100$ steps before switching to an iterative method. The hybrid method is compared with simulated annealing. Although the Metropolis-Hastings MCMC arrives at the steady state at about $10000$ Monte Carlo steps, in the hybrid method the Monte Carlo simulation can be stopped way before the burn-in time.
\end{abstract}

\maketitle

\section{Introduction}
\label{intro}

In near-infrared spectroscopy (NIRS), we estimate optical properties of biological tissue by solving inverse diffuse problems \cite{Boas-etal01,Hoshi11}. Such inverse problems are commonly solved by means of iterative methods. In the case of a homogeneous medium, absorption and reduced scattering coefficients of the medium can be obtained with iterative methods such as the Levenberg-Marquardt algorithm \cite{Levenberg44,Marquardt63} from time-resolved measurements (for example, see the review article \cite{Lange-Tachtsidis19}). Neuroimaging for the human brain by NIRS through the neurovascular coupling has been developed and is called functional NIRS (fNIRS)\cite{Chance-etal93,Hoshi-Tamura93,Kato-etal93,Villringer-etal93}. Since the region of interest on the head can be small, it is possible to assume a simple geometry of the half space. Quantitative measurements of inter-regional differences in neuronal activity requires accurate estimates of optical properties.

In Ref.~\cite{Laidevan-etal06}, the Nelder-Mead simplex method was used to retrieve optical parameters in layered tissue. Heterogeneity of optical properties can be obtained by diffuse optical tomography \cite{Boas-etal01,Gibson05,Arridege11}. Iterative numerical schemes are used to minimize the cost function when solving these inverse problems. A gradient-based approach \cite{Arridge-Schweiger98} was used to detect breast cancer \cite{Choe05}. The brain activity of a newborn infant was investigated \cite{Hebden-etal04} with diffuse optical tomography by TOAST (temporal optical absorption and scattering tomography) \cite{Arridge-Schweiger97,Schweiger-Arridge14}, in which iterative algorithms such as the nonlinear conjugate gradients, damped Gauss-Newton method, and Levenberg-Marquardt method are implemented. Diffuse optical tomography was performed on human lower legs and a forearm with the algebraic reconstruction algorithm in the framework of the modified generalized pulse spectrum technique \cite{Zhao-etal2005}. See Refs.~\cite{Arridge99,Arridge-Schotland09} for numerical techniques for diffuse optical tomography. For these iterative numerical schemes to work, it is important to choose a good initial guess for the initial value of the iteration.

Solving inverse problems by the Bayesian approach has been sought as an alternative way. In Ref.~\cite{Arridge-etal06}, a novel use of the Bayesian approach was considered to take modeling error into account. The Bayesian inversion with the Metropolis-Hastings Markov chain Monte Carlo was used in Refs.~\cite{Bal-Langmore-Marzouk13,Langmore-Davis-Bal13}. The Bayesian approach was used to determine optical parameters of the human head \cite{Bamett-Culver-Sorensen-Dale-Boas03}. Although the Markov chain Monte Carlo (MCMC) approach is in principle able to escape from local minima, it is computationally time consuming. Hence, despite the above-mentioned efforts, the use of Monte Carlo for inverse problems in NIRS has been extremely limited.

In this paper, we shed light on Markov chain Monte Carlo once again by combining it with an iterative scheme, and investigate the use of it in NIRS. In particular, we test a hybrid numerical scheme of Markov chain Monte Carlo and an iterative method. In the proposed hybrid scheme, the Markov chain Monte Carlo algorithm is first used to provide an initial guess using jumps in the landscape of the cost function, and then an iterative method is used after the initial large fluctuation. Thus the proposed method realizes a fast reconstruction while, in the beginning, obtained values at Monte Carlo steps jump around in the landscape of the cost function. The MCMC simulation is necessary only for the first $100$ steps. Then the hybrid method starts to use an iterative scheme and reconstructs optical properties by searching the global minimum. The computation time of the iterative scheme is negligible compared with that of the Monte Carlo simulation. Since the Monte Carlo simulation can be stopped way before the burn-in time for the hybrid method, the proposed scheme is at least ten times faster than simulated annealing, which is implemented by the naive use of the Metropolis-Hastings Markov chain Monte Carlo.

The rest of the paper is organized as follows. We develop diffusion theory in Sec.~2. A polyurethane-based phantom and numerical phantom are also described in Sec.~2. Section 3 is devoted to experimental and numerical results. Finally, discussion is given in Sec.~4.

\section{Materials and Methods}
\label{matemeth}

\subsection{Diffusion theory}
\label{de}

\subsubsection{Diffuse light in three dimensions}

Let us suppose that a random medium occupies the three-dimensional half space. We assume that the random medium is characterized by the diffusion coefficient $D$ and absorption coefficient $\mu_a$. We have $D=1/(3\mu_s')$, where $\mu_s'$ is the reduced scattering coefficient. Position in the half space ($-\infty<x<\infty$, $-\infty<y<\infty$, $0<z<\infty$) is denoted by $\vv{r}=(\boldsymbol{\rho},z)$, $\boldsymbol{\rho}=(x,y)$. Let $t$ denote time. Let $c$ be the speed of light in the medium. We assume an incident beam on the boundary at the origin in the $x$-$y$ plane. The energy density $u$ of light in the medium obeys the following diffusion equation.

\be
\left(\frac{1}{c}\frac{\pp}{\pp t}-D\Delta+\mu_a\right)u(\vv{r},t)=0,
\label{de3d}
\ee

\noindent
with the boundary condition

\be
-\ell\frac{\pp}{\pp z}u(\vv{r},t)+u(\vv{r},t)=\delta(x)\delta(y)q(t),
\ee

\noindent
and the initial condition $u(\vv{r},0)=0$. The right-hand side of the boundary condition is the incident laser beam which illuminates the medium at the origin $(0,0,0)$ with the temporal profile $q(t)$. In the phantom experiment described below, the incident light enters the phantom in the positive $z$ direction. The extrapolation distance $\ell$ is a nonnegative constant. If we consider the diffuse surface reflection, 
we have \cite{Groenhuis-Ferwerda-TenBosch83}

\be
\ell=2D\frac{1+r_d}{1-r_d},
\label{diffusereflection}
\ee

\noindent
where the internal reflection $r_d$ is given by \cite{Egan-Hilgeman79} $r_d=-1.4399n^{-2}+0.7099n^{-1}+0.6681+0.0636n$. Let us consider the corresponding surface Green's function $G_s(\vv{r},t;\boldsymbol{\rho}',s)$, which satisfies Eq.~(\ref{de3d}) and the boundary condition

\be
\left(-\ell\frac{\partial}{\partial z}+1\right)G_s(\vv{r},t;\boldsymbol{\rho}',s)=
\delta(x-x')\delta(y-y')\delta(t-s),
\ee

\noindent
together with the initial condition $G_s(\vv{r},0;\boldsymbol{\rho}',s)=0$. We obtain \cite{Carslaw-Jaeger59,Yosida-Ito,Hielscher-Jacques-Wang-Tittel95,MN17}

\be
\begin{aligned}
&
G_s(\vv{r},t;\boldsymbol{\rho}',s)=
\frac{cD}{\ell}\Biggl[\frac{2e^{-\mu_ac(t-s)}}{\left(4\pi Dc(t-s)\right)^{3/2}}
e^{-\frac{(x-x')^2+(y-y')^2}{4Dc(t-s)}}e^{-\frac{z^2}{4Dc(t-s)}}
\\
&-
\frac{e^{-\mu_ac(t-s)}}{4\pi\ell Dc(t-s)}
e^{-\frac{(x-x')^2+(y-y')^2}{4Dc(t-s)}}e^{\frac{z}{\ell}}
e^{\frac{Dc(t-s)}{\ell^2}}\mathop{\mathrm{erfc}}\left(
\frac{z+2Dc(t-s)/\ell}{\sqrt{4Dc(t-s)}}\right)\Biggr],
\quad t>s,
\end{aligned}
\ee

\noindent
with $G_s(\vv{r},t;\boldsymbol{\rho}',s)=0$ for $t\le s$. The complementary error function is defined by $\mathrm{erfc}(x)=(2/\sqrt{\pi})\int_x^{\infty}\exp(-t^2)\,dt$. 
Let $\vv{r}_d$ be a point in the $x$-$y$ plane with $|\vv{r}_d|>0$. We obtain

\be
\begin{aligned}
&
u(\vv{r}_d,t)=
\int_0^tG_s(\vv{r}_d,t;\vv{0},s)q(s)\,ds=
\int_0^te^{-\mu_ac(t-s)}e^{-\frac{|\vv{r}_d|^2}{4Dc(t-s)}}
\\
&\times
\frac{2q(s)}{\left(4\pi Dc(t-s)\right)^{3/2}}
\left[1-\frac{\sqrt{4\pi Dc(t-s)}}{2\ell}
e^{\frac{Dc(t-s)}{\ell^2}}
\mathop{\mathrm{erfc}}\left(
\frac{\sqrt{Dc(t-s)}}{\ell}\right)\right]\,ds.
\end{aligned}
\label{u3d}
\ee

\noindent
This $u(\vv{r}_d,t)$ in Eq.~(\ref{u3d}) is used for the calculation in Sec.~\ref{trs}.

\subsubsection{Diffuse light in two dimensions}

For the later purpose of a numerical experiment, we here consider light propagation in two dimensions. Let us suppose that the two-dimensional half space of positive $y$ is occupied with a random medium in the $x$-$y$ plane. The energy density $u$ of light at position $\boldsymbol{\rho}=(x,y)$ in the medium due to the incident beam on the boundary (the $x$-axis) at $\boldsymbol{\rho}_s^i$ ($i=1,\dots,M_s$) obeys the following diffusion equation.

\be
\left(\frac{1}{c}\frac{\pp}{\pp t}-D\Delta+\mu_a(\boldsymbol{\rho})\right)
u(\boldsymbol{\rho},t;\boldsymbol{\rho}_s^i)=0,
\label{de2d}
\ee

\noindent
with the boundary condition

\be
-\ell\frac{\pp}{\pp y}u(\boldsymbol{\rho},t;\boldsymbol{\rho}_s^i)+
u(\boldsymbol{\rho},t;\boldsymbol{\rho}_s^i)=
\delta(x-x_s^i)\delta(t),
\ee

\noindent
and the initial condition $u(\boldsymbol{\rho},0;\boldsymbol{\rho}_s^i)=0$. The incident laser beam on the right-hand side of the boundary condition was assumed to be a pulse at the position $\boldsymbol{\rho}_s^i=(x_s^i,0)$. We suppose that the diffusion coefficient $D$ is a positive constant but $\mu_a$ varies in space.

Below we develop the Rytov approximation. Let us write $\mu_a(\boldsymbol{\rho})\ge0$ as

\be
\mu_a(\boldsymbol{\rho})=\mu_{a0}+\delta\mu_a(\boldsymbol{\rho}),
\ee

\noindent
where $\mu_{a0}$ is a constant and the perturbation $\delta\mu_a(\boldsymbol{\rho})$ spatially varies. We note the relation,

\be
u(\boldsymbol{\rho},t;\boldsymbol{\rho}_s^i)=
u_0(\boldsymbol{\rho},t;\boldsymbol{\rho}_s^i)-
\int_0^t\int_0^{\infty}\int_{-\infty}^{\infty}
G(\boldsymbol{\rho},t;\boldsymbol{\rho}',s)
\delta\mu_a(\boldsymbol{\rho}')u(\boldsymbol{\rho}',s;\boldsymbol{\rho}_s^i)
\,dx'dy'ds,
\label{formalsol}
\ee

\noindent
where $u_0(\boldsymbol{\rho},t;\boldsymbol{\rho}_s^i)$ is the solution to the diffusion equation (\ref{de2d}) with the zeroth-order coefficient replaced by $b_0$. We introduce the Green's function $G(\boldsymbol{\rho},t;\boldsymbol{\rho}',s)$, which satisfies

\be
\left(\frac{1}{c}\frac{\pp}{\pp t}-D\Delta+\mu_{a0}\right)G=
\delta(\boldsymbol{\rho}-\boldsymbol{\rho}')\delta(t-s),
\ee

\noindent
with the boundary condition $-\ell\frac{\pp G}{\pp y}+G=0$ at $y=0$, and the initial condition $G=0$ at $t=0$. The above relation (\ref{formalsol}) can be directly verified. By recursive substitution, we obtain $u$ as

\be
\begin{aligned}
u(\boldsymbol{\rho},t;\boldsymbol{\rho}_s^i)
&=
u_0(\boldsymbol{\rho},t;\boldsymbol{\rho}_s^i)-
\int_0^t\int_0^{\infty}\int_{-\infty}^{\infty}
G(\boldsymbol{\rho},t;\boldsymbol{\rho}',s)\delta\mu_a(\boldsymbol{\rho}')
u_0(\boldsymbol{\rho}',s;\boldsymbol{\rho}_s^i)\,dx'dy'ds
\\
&+
O((\delta\mu_a)^2).
\end{aligned}
\ee

\noindent
By neglecting higher-order terms assuming that $\delta\mu_a$ is small, we arrive at the (first) Born approximation \cite{Ishimaru78}, in which $u$ is given by

\be
u(\boldsymbol{\rho},t;\boldsymbol{\rho}_s^i)=
u_0(\boldsymbol{\rho},t;\boldsymbol{\rho}_s^i)-
c\int_0^t\int_0^{\infty}\int_{-\infty}^{\infty}
G(\boldsymbol{\rho},t;\boldsymbol{\rho}',s)\delta\mu_a(\boldsymbol{\rho}')
u_0(\boldsymbol{\rho}',s;\boldsymbol{\rho}_s^i)\,dx'dy'ds.
\ee

We note that the Green's function is obtained as \cite{Carslaw-Jaeger59,Yosida-Ito,Hielscher-Jacques-Wang-Tittel95,MN17}

\be
\begin{aligned}
&
G(\boldsymbol{\rho},t;\boldsymbol{\rho}',s)=
\frac{e^{-\mu_{a0}c(t-s)}}{4\pi D(t-s)}e^{-\frac{(x-x')^2}{4Dc(t-s)}}
\Biggl[e^{-\frac{(y-y')^2}{4Dc(t-s)}}+e^{-\frac{(y+y')^2}{4Dc(t-s)}}
\\
&-
\frac{\sqrt{4\pi Dc(t-s)}}{\ell}e^{-\frac{(y+y')^2}{4Dc(t-s)}}
e^{\left(\frac{y+y'+2Dc(t-s)/\ell}{\sqrt{4Dc(t-s)}}\right)^2}
\mathrm{erfc}\left(\frac{y+y'+2Dc(t-s)/\ell}{\sqrt{4Dc(t-s)}}\right)
\Biggr],
\end{aligned}
\ee

\noindent
for $t>s$, and otherwise $G(\boldsymbol{\rho},t;\boldsymbol{\rho}',s)=0$. Moreover we obtain

\be
u_0(\boldsymbol{\rho},t;\boldsymbol{\rho}_s^i)=
\frac{e^{-\mu_{a0}ct}}{2\pi Dt}e^{-\frac{(x-x_s^i)^2+y^2}{4Dct}}
\left[1-
\frac{\sqrt{\pi Dct}}{\ell}e^{\left(\frac{y+2Dct/\ell}{\sqrt{4Dct}}\right)^2}
\mathrm{erfc}\left(\frac{y+2Dct/\ell}{\sqrt{4Dct}}\right)\right].
\ee

\noindent
The above expression of $u_0$ is similar to the formula in Eq.~(\ref{u3d}) but does not have a time integral because in this case we assumed the delta function $\delta(t)$ for the incident beam.

To obtain the expression of the Rytov approximation, we introduce $\psi_0,\psi_1$ as \cite{Ishimaru78}

\be
u_0=e^{\psi_0},\qquad u=e^{\psi_0+\psi_1}.
\ee

\noindent
By plugging the above expressions of $u,u_0$ into the Born approximation and neglecting terms of $O((\delta\mu_a)^2)$, we obtain

\be
\begin{aligned}
e^{\psi_1}
&=
1-e^{-\psi_0}\int_0^t\int_0^{\infty}\int_{-\infty}^{\infty}
G(\boldsymbol{\rho},t;\boldsymbol{\rho}',s)\delta\mu_a(\boldsymbol{\rho}')
u_0(\boldsymbol{\rho}',s;\boldsymbol{\rho}_s^i)\,dx'dy'ds
\\
&=
\exp\left[-e^{-\psi_0}\int_0^t\int_0^{\infty}\int_{-\infty}^{\infty}
G(\boldsymbol{\rho},t;\boldsymbol{\rho}',s)\delta\mu_a(\boldsymbol{\rho}')
u_0(\boldsymbol{\rho}',s;\boldsymbol{\rho}_s^i)\,dx'dy'ds
\right].
\end{aligned}
\ee

\noindent
The (first) Rytov approximation is thus derived as

\be
\begin{aligned}
&
u(\boldsymbol{\rho},t;\boldsymbol{\rho}_s^i)=
u_0(\boldsymbol{\rho},t;\boldsymbol{\rho}_s^i)
\\
&\times
\exp\Biggl[-\frac{1}{u_0(\boldsymbol{\rho},t;\boldsymbol{\rho}_s^i)}
\int_0^t\int_0^{\infty}\int_{-\infty}^{\infty}
G(\boldsymbol{\rho},t;\boldsymbol{\rho}',s)\delta\mu_a(\boldsymbol{\rho}')
u_0(\boldsymbol{\rho}',s;\boldsymbol{\rho}_s^i)\,dx'dy'ds\Biggr].
\end{aligned}
\label{rytov:rytov1}
\ee

Let us define

\be
\begin{aligned}
g(y,t;y',s)&=
\frac{1}{4\pi D(t-s)}e^{-\frac{(y+y')^2}{4Dc(t-s)}}
\Biggl[1+e^{\frac{(y+y')^2-(y-y')^2}{4Dc(t-s)}}-
\frac{\sqrt{4\pi Dc(t-s)}}{\ell}
\\
&\times
e^{\left(\frac{y+y'}{2\sqrt{Dc(t-s)}}+\frac{\sqrt{Dc(t-s)}}{\ell}\right)^2}
\mathrm{erfc}\left(\frac{y+y'}{2\sqrt{Dc(t-s)}}+\frac{\sqrt{Dc(t-s)}}{\ell}
\right)
\Biggr].
\end{aligned}
\ee

\noindent
Then we have

\be
\begin{aligned}
&
\int_0^tG(\boldsymbol{\rho},t;\boldsymbol{\rho}',s)u_0(\boldsymbol{\rho}',s)
\,ds
\\
&=
e^{-\mu_{a0}ct}\int_0^te^{-\frac{(x-x')^2}{4Dc(t-s)}}
e^{-\frac{(x'-x_s^i)^2}{4Dcs}}g(y,t;y',s)g(y',s;0,0)\,ds.
\end{aligned}
\ee

\noindent
Therefore, Eq.~(\ref{rytov:rytov1}) can be rewritten as

\be
\begin{aligned}
u(\boldsymbol{\rho},t;\boldsymbol{\rho}_s^i)
&=
u_0(\boldsymbol{\rho},t;\boldsymbol{\rho}_s^i)
\exp\Bigg[-
\frac{e^{-\mu_{a0}ct}}{u_0(\boldsymbol{\rho},t;\boldsymbol{\rho}_s^i)}
\int_0^{\infty}\int_{-\infty}^{\infty}\delta\mu_a(\boldsymbol{\rho}')
\\
&\times
\left(\int_0^te^{-\frac{(x-x')^2}{4Dc(t-s)}}
e^{-\frac{(x'-x_s^i)^2}{4Dcs}}g(y,t;y',s)g(y',s;0,0)\,ds\right)\,dx'dy'
\Bigg].
\end{aligned}
\label{fwd:ufinal}
\ee

\noindent
This expression (\ref{fwd:ufinal}) is used to compute the forward data in Sec.~\ref{toy}.

\subsection{Inverse problems by an iterative scheme}
\label{LM}

We suppose that on the surface of a two- or three-dimensional random medium, for each source at $\boldsymbol{\rho}_s^i$ or $\vv{r}_s^i$ ($i=1,\dots,M_s$) exiting light is detected at $\boldsymbol{\rho}_d^j$ or $\vv{r}_d^j$ ($j=1,\dots,M_d$) and is measured at times $t^k$ ($k=1,\dots,M_t$). Let $y_{ijk}$ be measured data whereas $u$ is a solution to the diffusion equation. Let us suppose that $u$ depends on a vector $\vv{a}$ which contains unknown parameters. We wish to reconstruct $\vv{a}$. In Sec.~\ref{trs}, $\vv{a}=(\mu_a,D)$, and $a$ is a scalar (a one-dimensional vector) in Sec.~\ref{toy}. For example, in the former case the solution $u$ depends on $\vv{a}$ since the calculated value of $u$ depends on $\mu_a,D$.

Let us introduce vectors $\vv{U}$ and $\vv{F}(\vv{a})$ of dimension $M_sM_dM_t$ as

\be
\{\vv{U}\}_{ijk}=y_{ijk},\qquad \{\vv{F}(\vv{a})\}_{ijk}=
u(\boldsymbol{\rho}_d^j,t^k;\boldsymbol{\rho}_s^i;\vv{a})\;\mbox{or}\;
u(\vv{r}_d^j,t^k;\vv{r}_s^i;\vv{a}),
\ee

\noindent
where we wrote $u(\boldsymbol{\rho}_d^j,t^k;\boldsymbol{\rho}_s^i)=u(\boldsymbol{\rho}_d^j,t^k;\boldsymbol{\rho}_s^i;\vv{a})$ and $u(\vv{r}_d^j,t^k;\vv{r}_s^i)=u(\vv{r}_d^j,t^k;\vv{r}_s^i;\vv{a})$ emphasizing that $u$ depends on $\vv{a}$. We find optimal $\vv{a}$ by minimizing $\|\vv{U}-\vv{F}(\vv{a})\|_2^2$. Here we particularly consider the Levenberg-Marquardt method \cite{Levenberg44,Marquardt63}. That is, the reconstructed value $\vv{a}^*=\lim_{k\to\infty}\vv{a}_k$ is computed by the iteration given by

\be
\vv{a}_{k+1}=\vv{a}_k+\left[F'(\vv{a}_k)^TF'(\vv{a}_k)+\lambda I\right]^{-1}F'(\vv{a}_k)^T\left(\vv{U}-\vv{F}(\vv{a}_k)\right),
\label{LM:iteration}
\ee

\noindent
where $F'(\vv{a})$ is the Jacobian matrix, which contains derivatives of $\vv{F}(\vv{a})$ with respect to $\vv{a}$, and the parameter $\lambda$ is nonnegative. By modifying the original algorithm according to Ref.~\cite{Fletcher71}, our algorithm of the Levenberg-Marquardt method, which we call Algorithm 1, is described below.

\vspace{1em}
\paragraph*{\bf Algorithm 1: Levenberg-Marquardt (LM)}
\begin{enumerate}
\item Set $k=0$ and $\lambda=1$.
\item Calculate $\vv{F}(\vv{a}_k)$ and $F'(\vv{a}_k)$.
\item Calculate $S(\vv{a}_k)=\vv{d}^T\vv{d}$, where $\vv{d}=\vv{U}-\vv{F}(\vv{a}_k)$.
\item Prepare $A=F'(\vv{a}_k)^TF'(\vv{a}_k)$ and $\vv{v}=F'(\vv{a}_k)^T\vv{d}$.
\item Find $\boldsymbol{\delta}$ from $(A+\lambda I)\boldsymbol{\delta}=-\vv{v}$.
\item Obtain $S(\vv{a}_k+\boldsymbol{\delta})$ and $R=\frac{S(\vv{a}_k)-S(\vv{a}_k+\boldsymbol{\delta})}{-\boldsymbol{\delta}^T(2\vv{v}+A\boldsymbol{\delta})}$.
\item If $R<0.25$, then set $\nu=10$ ($\alpha_c<0.1$), $\nu=1/\alpha_c$ $(0.1\le\alpha_c\le0.5)$, or $\nu=2$ ($\alpha_c>0.5$), where $\alpha_c=\left[2-\left(S(\vv{a}_k+\boldsymbol{\delta})-S(\vv{a}_k)\right)/\boldsymbol{\delta}^T\vv{v}\right]^{-1}$. If $R<0.25$ and $\lambda=0$, set $\lambda=1/\|A^{-1}\|$ and $\nu=\nu/2$. In the case of $R<0.25$, we set $\lambda=\nu\lambda$. If $R>0.75$, then we set $\lambda=\lambda/2$. If $R>0.75$ and $\lambda<1/\|A^{-1}\|$, set $\lambda=0$. Otherwise when $0.25\le R\le0.75$, no update for $\lambda$.
\item If $S(\vv{a}_k+\boldsymbol{\delta})\ge S(\vv{a}_k)$, then return to Step 5.
\item If $S(\vv{a}_k+\boldsymbol{\delta})<S(\vv{a}_k)$, set $\vv{a}_{k+1}=\vv{a}_k+\boldsymbol{\delta}$. Then put $k+1\rightarrow k$ and go back to Step 2. Repeat the above procedure until one of the stopping criteria $\|\boldsymbol{\delta}\|<10^{-4}$ and $S<10^{-14}$ is fulfilled.
\end{enumerate}

\subsection{Inverse problems by Markov chain Monte Carlo}
\label{MCMC}

For simplicity in this section, we describe the Metropolis-Hastings Markov chain Monte Carlo algorithm (MH-MCMC) assuming $a$ is the scalar appearing in Sec.~\ref{toy}. The extension of the derivation to the general case of vector $\vv{a}$ is straightforward.

Suppose that the coefficient $\mu_a$ is unknown and $\mu_a$ depends on a parameter $a$. We will reconstruct $a$ within the framework of the Bayesian inversion algorithm \cite{Kaipio-Somersalo,Nakamura-Potthast}. That is, we will find the probability distribution $\pi(a|\vv{U})$ of $a$ for measured data $\vv{U}$. Let $f_{\rm prior}(a)$ be the prior probability density and $\pi(\vv{U}|a)$ be the likelihood density or the conditional probability density of $\vv{U}$ for $a$. Then the joint probability density $\pi(a,\vv{U})$ of $a,\vv{U}$ is given by $\pi(a,\vv{U})=\pi(\vv{U}|a)f_{\rm prior}(a)$. According to the Bayes formula, the conditional probability density $\pi(a|\vv{U})$ is given by

\be
\pi(a|\vv{U})=
\frac{L(\vv{U}|a)f_{\rm prior}(a)}
{\int_{-\infty}^{\infty}L(\vv{U}|a)f_{\rm prior}(a)\,da},
\ee

\noindent
where $L(\vv{U}|a)$ is a function proportional to $\pi(\vv{U}|a)$. Assuming Gaussian noise, we put

\be
L(\vv{U}|a)=e^{-\frac{1}{2\sigma_e^2}\|\vv{U}-\vv{F}(a)\|_2^2}.
\ee

\noindent
In this paper we simply set $f_{\rm prior}(a)=1$. That is, we can say $f_{\rm prior}(a)\propto{\bf 1}_{[a_{\rm min},a_{\rm max}]}(a)$ and the interval $[a_{\rm min},a_{\rm max}]$ is large enough so that all $a$'s appearing in the Markov chain fall into this interval. Here, ${\bf 1}_A(a)$ is the indicator function defined as ${\bf 1}_A(a)=1$ for $a\in A$ and $=0$ for $a\notin A$. General uniform distributions can be used for $f_{\rm prior}(a)$ if we use the prior-reversible proposal that satisfies $f_{\rm prior}(a)q(a'|a)=f_{\rm prior}(a')q(a|a')$ (see below for the proposal distribution $q(a'|a)$) \cite{Iglesias-Lin-Stuart14}. Another possible choice of $f_{\rm prior}(a)$ is a Gaussian distribution, which turns out to be the Tikhonov regularization term in the cost function.

Using the Metropolis-Hastings algorithm, we can evaluate $\pi(a|\vv{U})$ even when the normalization factor is not known and only the relation $\pi(a|\vv{U})\propto L(\vv{U}|a)f_{\rm prior}(a)$ is available \cite{Kaipio-Somersalo,Nakamura-Potthast}. We can find $\pi(a|\vv{U})$ using a sequence $p_1,p_2,\dots$ as

\be
\pi(a|\vv{U})=\lim_{k\to\infty}p_k(a),
\ee

\noindent
where $p_{k+1}(a)$ is obtained from $p_k(a)$ (Markov chain) as

\be
p_{k+1}(a')=\int_{-\infty}^{\infty}K(a',a)p_k(a)\,da.
\ee

\noindent
For all $a,a'$, the transition kernel satisfies

\be
K(a',a)\ge0,\qquad\int_{-\infty}^{\infty}K(a',a)\,da'=1.
\ee

\noindent
Let us write $K(a',a)$ as

\be
K(a',a)=\alpha(a',a)q(a'|a)+r(a)\delta(a'-a),
\ee

\noindent
where $q(a'|a)$ is the proposal distribution and

\be
r(a)=1-\int_{-\infty}^{\infty}\alpha(a',a)q(a'|a)\,da'.
\ee

\noindent
In the Metropolis-Hastings algorithm we set $\alpha(a',a)=\min\left\{1,\pi(a'|\vv{U})q(a|a')/[\pi(a|\vv{U})q(a'|a)]\right\}$. With this choice of $\alpha(a',a)$, the detailed balance is satisfied and $K(a,a')\pi(a'|\vv{U})=K(a',a)\pi(a|\vv{U})$. A common choice of $q(a'|a)$ is the normal distribution, i.e., $q(\cdot|a)=\mathcal{N}(a,\varepsilon^2)$ with the mean $a$ and the standard deviation $\varepsilon>0$. We note that $q(a'|a)=q(a|a')$ in this case. We have

\be
\int_{-\infty}^{\infty}h(a)\pi(a|\vv{U})\,da=
\lim_{k_{\rm max}\to\infty}\frac{1}{k_{\rm max}}
\sum_{k=1}^{k_{\rm max}}h(a_k),
\ee

\noindent
where $a_k\sim p_k(\cdot)$ and $h$ is an arbitrary continuous bounded function.

Simulated annealing \cite{Kirkpatrick-GelattJr-Vecchi83} is a type of the Metropolis-Hastings MCMC algorithm in which the {\em temperature} $\sigma_e$ decreases during the simulation. The algorithm is summarized below as Algorithm 2. In this paper we consider two temperatures.

\vspace{1em}
\paragraph*{\bf Algorithm 2: Two-temperature simulated annealing (SA)}
\begin{enumerate}
\item Set large $\sigma_e$ as a high temperature.
\item Generate $a'\sim q(\cdot|a_k)=\mathcal{N}(a_k,\varepsilon^2)$ 
with $\varepsilon>0$ for given $a_k$.
\item Calculate $\alpha(a',a_k)=\min\left\{1,\pi(a'|\vv{U})/\pi(a_k|\vv{U})\right\}$.
\item Update $a_k$ as $a_{k+1}=a'$ with probability $\alpha(a',a_k)$ but 
otherwise set $a_{k+1}=a_k$.
\item Continue while $k\le k_b$.
\item Then decrease $\sigma_e$ to a smaller value as a low temperature, and continue to update $a_k$.
\end{enumerate}

Now we propose an MCMC-iterative hybrid method by combining Algorithms 1 with the Metropolis-Hastings Markov chain Monte Carlo. We first run the Markov chain Monte Carlo. Although the Metropolis-Hastings MCMC eventually gives the correct solution after about $10000$ steps when the chain reaches the steady state, we stop the Monte Carlo simulation way before the burn-in time. At the $k_b$th Monte Carlo step after initial rapid changes ceases, we record the obtained reconstructed value $a_{k_b}$ and switch to Algorithm 1 with the initial value the recorded $a_{k_b}$. It is important that $k_b$ is chosen after the initial rapid changes although $k_b$ can be significantly a way from the burn-in time. Otherwise, the final reconstructed results are quite robust against the choice of $k_b$. We refer to the following hybrid algorithm as Algorithm 3.

\vspace{1em}
\paragraph*{\bf Algorithm 3: Hybrid}
\begin{enumerate}
\item Choose an initial guess $a_0$, which is not necessarily close to 
the global minimum. 
\item Generate $a'\sim q(\cdot|a_k)=\mathcal{N}(a_k,\varepsilon^2)$ 
with $\varepsilon>0$ for given $a_k$.
\item Calculate $\alpha(a',a_k)=\min\left\{1,\pi(a'|U)/\pi(a_k|U)\right\}$.
\item Update $a_k$ as $a_{k+1}=a'$ with probability $\alpha(a',a_k)$ but 
otherwise set $a_{k+1}=a_k$.
\item Obtain a reconstructed $a_{k_b}$.
\item Switch to Algorithm 1 with the initial guess $a_{k_b}$.
\end{enumerate}

We close this subsection by the Gelman-Rubin convergence diagnostic, which uses the intra-chain variance and inter-chain variance \cite{Gelman-Rubin92,Martinez-Martinez15}. We run $M_{\rm MC}$ different chains with different initial values. Let $a_k^m$ denote the $k$th value in the $m$th chain ($k=1,\dots,k_b$, $m=1,\dots,M_{\rm MC}$). We discard the first $k_a-1$ steps before the initial rapid changes cease. Then we compute the following intra-chain average and variance.

\be
\bar{a}_m=\frac{1}{k_b-k_a+1}\sum_{k=k_a}^{k_b}a_k^m,\quad
\sigma_m^2=\frac{1}{k_b-k_a}\sum_{k=k_a}^{k_b}(a_k^m-\bar{a}_m)^2.
\ee

\noindent
Next we compute the inter-chain mean and variance.

\be
\bar{a}=\frac{1}{M_{\rm MC}}\sum_{m=1}^{M_{\rm MC}}\bar{a}_m,\quad
B=\frac{k_b-k_a+1}{M_{\rm MC}-1}\sum_{m=1}^{M_{\rm MC}}(\bar{a}_m-\bar{a})^2.
\ee

\noindent
Now we introduce

\be
\hat{V}=\frac{k_b-k_a}{k_b-k_a+1}W+\frac{1}{k_b-k_a+1}B,
\ee

\noindent
where $W=\frac{1}{M_{\rm MC}}\sum_{m=1}^{M_{\rm MC}}\sigma_m^2$. This is an unbiased estimator of the true variance. But $W$ is also an unbiased estimate of the true variance if the chains converge. Therefore we have $\sqrt{\hat{V}/W}\approx1$ if converged. Below, we will see that we can choose $k_b$ for which $\sqrt{\hat{V}/W}$ is not close to $1$. In this paper, we set $M_{\rm MC}=10$.

\subsection{TRS measurements of a polyurethane-based phantom}

\begin{figure}[ht]
\centering
\includegraphics[height=6cm]{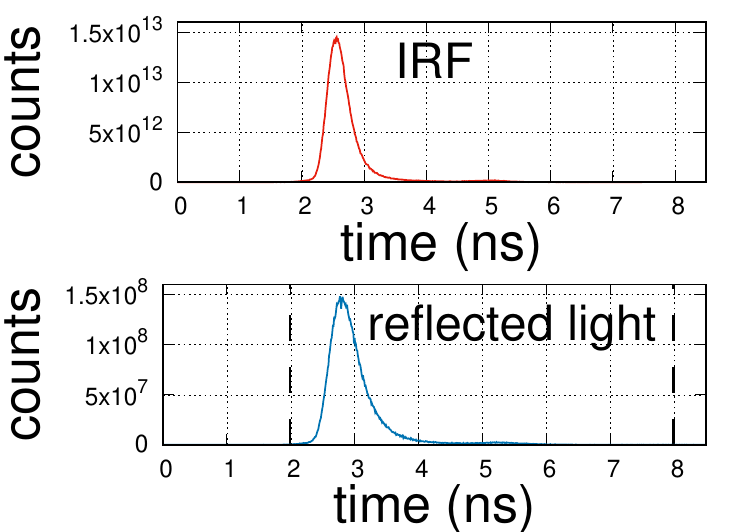}
\caption{\label{trs:trs}
Measured data from TRS-80. The instrument response function and measured reflected light are shown in the upper and lower panels, respectively. In the lower panel, the dashed lines show $t_1=2.00\,{\rm ns}$ and $t_{M_t}=8.00\,{\rm ns}$ ($M_t=601$).
}
\end{figure}

In this section, we consider time-resolved measurements for a phantom. The solid phantom (biomimic optical phantom) is made of polyurethane to simulate biological tissues (INO, Quebec, QC, Canada). The absorption coefficient and reduced scattering coefficient are $\mu_a=0.0209\,{\rm mm}^{-1}$ and $\mu_s'=0.853\,{\rm mm}^{-1}$ at wavelength 800 nm. The refractive index of the phantom is $n=1.51$. The measurements were performed by the TRS (time-resolved spectroscopy) instrument (TRS-80, Hamamatsu Photonics K.K., Hamamatsu, Japan). Two optical fibers from TRS-80 are attached on the top of the phantom with the separation $13\,{\rm mm}$ ($M_s=M_d=1$). The phantom is illuminated by near-infrared light of the wavelength $760\,{\rm nm}$ through one optical fiber and the reflected light is detected by the other optical fiber. The time interval of the measured data is $10\,{\rm ps}$ and we used the data from $t_1=2.00\,{\rm ns}$ to $t_{M_t}=8.00\,{\rm ns}$ ($M_t=601$). Measured counts, i.e., the number of photons, are shown in Fig.~\ref{trs:trs}. The upper panel of Fig.~\ref{trs:trs} shows the instrument response function (IRF), which is given as a property of the experimental device. The measured reflected light is shown in the lower panel of Fig.~\ref{trs:trs}. The function $q(t^k)$ is the instrument response function divided by the maximum value of the measured reflected light. The peak of $q$ is at $2.56\,{\rm ns}$. $\{U\}_{11k}=y_{11k}$ is the measured reflected light normalized by its maximum value. In the lower panel of Fig.~\ref{trs:trs}, $t_1$ and $t_{M_t}$ are marked by vertical dashed lines. The size of the phantom is large enough that we can assume the three-dimensional half space. Then the energy density of the detected light is computed by Eq.~(\ref{u3d}). The parameter $\ell$ is obtained from Eq.~(\ref{diffusereflection}). 

\subsection{Numerical phantom}

In order to examine when the iterative scheme fails by being trapped by a local minimum and how the Markov chain Monte Carlo is capable of escaping from such local minima, a toy model is devised which is simple enough to explicitly understand the structure of the cost function but is complicated enough that the cost function has one local minimum and one global minimum.

\begin{figure}[ht]
\centering
\includegraphics[height=6cm]{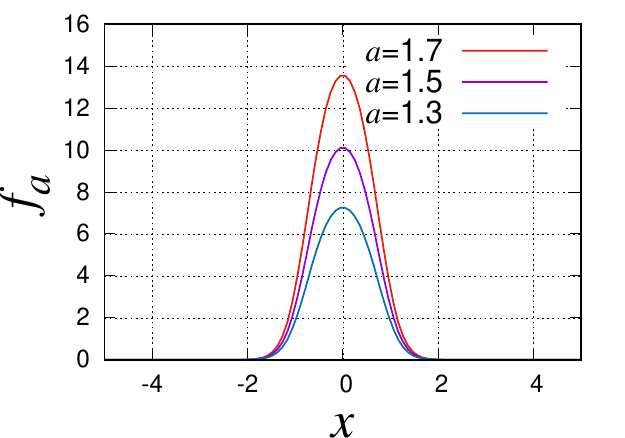}
\caption{\label{fwd:figfa}
The function $f_a(x)$ in Eq.~(\ref{fwd:delaf}) is plotted for $a=1.3$, $1.5$, and $1.7$.
}
\end{figure}

We consider diffuse optical tomography in the two-dimensional space. In our toy model we suppose that the diffusion coefficient $D$ is constant everywhere in the medium but there is absorption inhomogeneity and the absorption coefficient $\delta\mu_a$ is unknown. Moreover we assume that $\delta\mu_a(\boldsymbol{\rho})$ is given by

\be
\delta\mu_a(\boldsymbol{\rho})=\eta f_a(x)\delta(y-y_0),
\label{fwd:dela}
\ee

\noindent
where $\eta,y_0$ are given positive constants. Here, $f_a(x)$ is given by

\be
f_a(x)=\left[a^3+3\left(1+\frac{\tanh{x^2}}{10}\right)a^2\right]
\left(1-\tanh{x^2}\right),
\label{fwd:delaf}
\ee

\noindent
where $a>1.1$ is a constant. Thus, $a$ is the parameter to be reconstructed in our toy inverse problem. As is shown in Fig.~\ref{fwd:figfa}, the function $f_a(x)$ has one peak at $x=0$ and the maximum value is $f_a(0)=a^2(a+3)$, $f_a$ monotonically decays for $|x|>0$, and $f_a\to0$ as $|x|\to\infty$.

Now we describe how the forward data is computed. The unit of length and unit of time are taken to be ${\rm mm}$ and ${\rm ps}$, respectively. On the $x$-axis, we place two sources ($M_s=2$) at $\boldsymbol{\rho}_s^1=(-20,0)$, $\boldsymbol{\rho}_s^2=(20,0)$ and three detectors ($M_d=3$) at $\boldsymbol{\rho}_d^1=(-40,0)$, $\boldsymbol{\rho}_d^2=(0,0)$, $\boldsymbol{\rho}_d^3=(40,0)$. We set $\mu_s'=1$, $\mu_a=0.02$, $n=1.37$. Suppose that there is absorption inhomogeneity at depth $5$. For $\delta\mu_a$, we put $\eta=300/c$, $y_0=5$, and

\be
a=1.5.
\ee

\noindent
To distinguish, hereafter the true value of $a$ is denoted by $\bar{a}$. We assume $3\%$ Gaussian noise and give the measured data as

\be
y_{ijk}=u(\boldsymbol{\rho}_d^j,t^k;\boldsymbol{\rho}_s^i;\bar{a})(1+e_{ijk}),
\ee

\noindent
where $e_{ijk}\sim\mathcal{N}(0,0.03^2)$. In this numerical experiment, $M_t=500$ and $t^{k+1}-t^k=t^1=5$ ($k=1,\dots,M_t-1$).

By substituting the form (\ref{fwd:dela}) of $\delta b$ in 
(\ref{fwd:ufinal}), we can express the energy density as

\be
\begin{aligned}
u(\boldsymbol{\rho}_d^j,t^k;\boldsymbol{\rho}_s^i;a)
&=
u(\boldsymbol{\rho}_d^j,t^k;\boldsymbol{\rho}_s^i)
\\
&=
u_0(\boldsymbol{\rho}_d^j,t^k;\boldsymbol{\rho}_s^i)
\exp\Bigg[-\frac{\eta e^{-\mu_{a0}ct^k}}{u_0(\boldsymbol{\rho}_d^j,t^k;\boldsymbol{\rho}_s^i)}\int_0^{t^k}g(0,t^k;y_0,s)
\\
&\times
g(y_0,s;0,0)\left(\int_{-\infty}^{\infty}f_a(x')
e^{-\frac{(x_d^j-x')^2}{4Dc(t^k-s)}}e^{-\frac{(x'-x_s^i)^2}{4Dcs}}
\,dx'\right)\,ds\Bigg],
\end{aligned}
\label{fwd:formula}
\ee

\noindent
where

\be
u_0=\frac{e^{-\mu_{a0}ct^k}}{2\pi Dt^k}e^{-\frac{(x_d^j-x_s^i)^2}{4Dct^k}}
\left[1-\frac{\sqrt{\pi Dct^k}}{\ell}
e^{\left(\frac{\sqrt{Dct^k}}{\ell}\right)^2}
\mathrm{erfc}\left(\frac{\sqrt{Dct^k}}{\ell}\right)\right].
\ee

\section{Results}

\subsection{Determination of optical properties}
\label{trs}

Here we consider reconstructions from measured data in the phantom experiment. Let us first consider initial guesses below.

\be
\mbox{Case 1:}\qquad
\mu_a=0.01\,{\rm mm}^{-1},\quad\mu_s'=1.0\,{\rm mm}^{-1}.
\ee

\noindent
In this Case 1, the following $\mu_a,\mu_s'$ are obtained both by Algorithm 1 (LM) and Algorithm 3 (hybrid).

\be
\mu_a=0.016\,{\rm mm}^{-1},\quad\mu_s'=0.63\,{\rm mm}^{-1}.
\label{correctanswer}
\ee

\begin{figure}[ht]
\centering
\includegraphics[height=6cm]{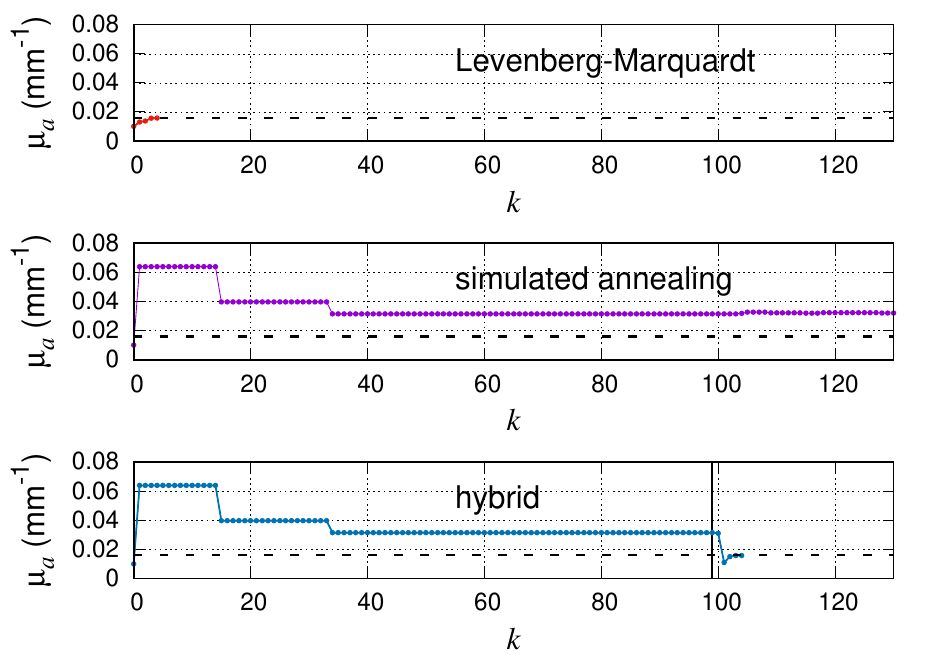}
\caption{\label{trs:fig1}
Case 1: Reconstructed $\mu_a$ by (top) Algorithm 1, (middle) Algorithm 2, and (bottom) Algorithm 3. The dashed lines show $\mu_a$ in Eq.~(\ref{correctanswer}).
}
\end{figure}

\begin{figure}[ht]
\centering
\includegraphics[height=6cm]{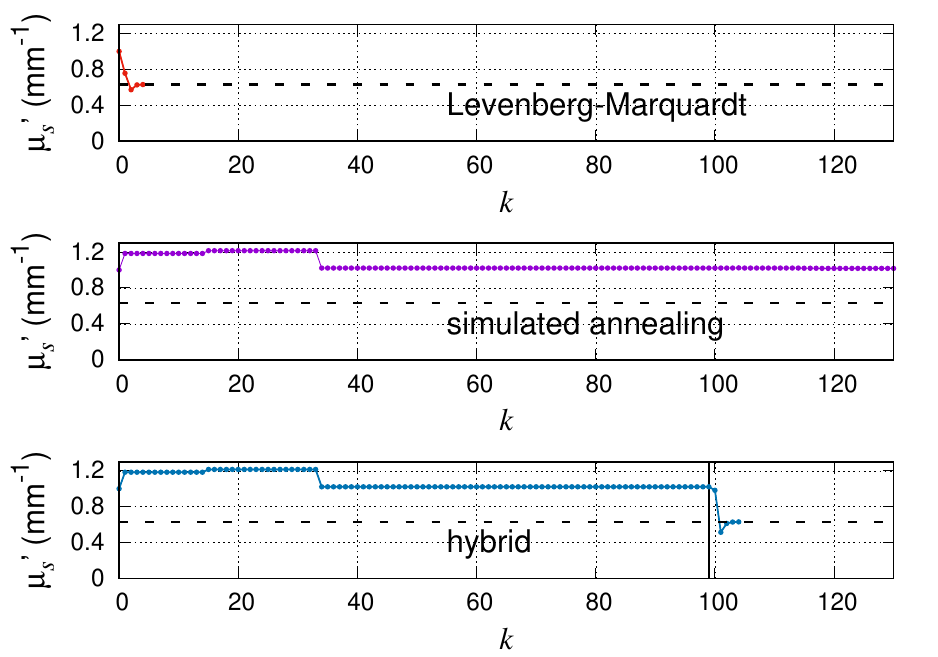}
\caption{\label{trs:fig2}
Case 1: Reconstructed $\mu_s'$ by (top) Algorithm 1, (middle) Algorithm 2, and (bottom) Algorithm 3. The dashed lines show $\mu_s'$ in Eq.~(\ref{correctanswer}).
}
\end{figure}

\noindent
The results for $\mu_a$ and $\mu_s'$ are shown in Figs.~\ref{trs:fig1} and \ref{trs:fig2}, respectively. In Fig.~\ref{trs:fig1}, reconstructed values of $\mu_a$ are plotted against the iteration number $k$. In the top panel of Fig.~\ref{trs:fig1}, we see that Algorithm 1 (LM) quickly converges. In Algorithms 2 and 3, we set $k_b=99$. As is seen in the middle panel of Fig.~\ref{trs:fig1}, the convergence of Algorithm 2 (SA) is slow. The {\em temperature} is decreased from $\sigma_e=10^{-6}$ to $\sigma_e=10^{-7}$ at the $k_b$th Monte Carlo step. Similarly, $\varepsilon$ is changed from $0.1$ to $0.001$. Algorithm 2 returns the correct values after many Monte Carlo steps; we have $\sqrt{\hat{V}/W}=1.02,1.06$ for $\mu_a,\mu_s'$, respectively, when $k_a=9000$ and $k_b=10000$. Finally in the bottom panel of Fig.~\ref{trs:fig1}, we see that the iteration of Algorithm 3 (hybrid) immediately converges after we switch from the Metropolis-Hastings MCMC ($\sigma_e=10^{-6}$, $\varepsilon=0.1$) to Algorithm 1 (LM). No convergence of the Monte Carlo chain is required, and we found $\sqrt{\hat{V}/W}=8.6$ for $\mu_a$ and $=11.0$ for $\mu_s'$ ($k_a=49$, $k_b=99$). A similar behavior is observed for the reconstruction of $\mu_s'$ in Fig.~\ref{trs:fig2}. In the top panel of Fig.~\ref{trs:fig2}, we find that Algorithm 1 (LM) works best and converges after a few iterations, whereas Algorithm 2 (SA) in the middle panel of Fig.~\ref{trs:fig2} has slow convergence and reconstructed values around at $k_b=99$ are still away from $\mu_s'$ in (\ref{correctanswer}). By switching from MH-MCMC to Algorithm 1 (LM) using Algorithm 3 (hybrid), convergence is easily obtained as shown in the bottom panel of Fig.~\ref{trs:fig2}.

Now we start the simulation by setting the following initial values.

\be
\mbox{Case 2:}\qquad
\mu_a=0.5\,{\rm mm}^{-1},\quad\mu_s'=1.0\,{\rm mm}^{-1}.
\ee

\begin{figure}[ht]
\centering
\includegraphics[height=4.5cm]{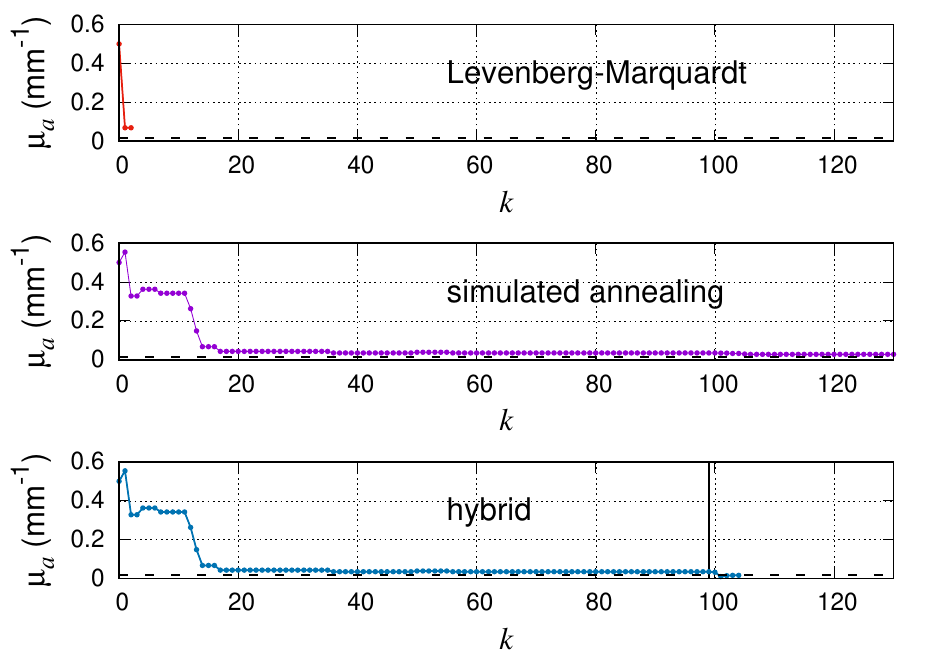}\hspace{-4mm}
\includegraphics[height=4.5cm]{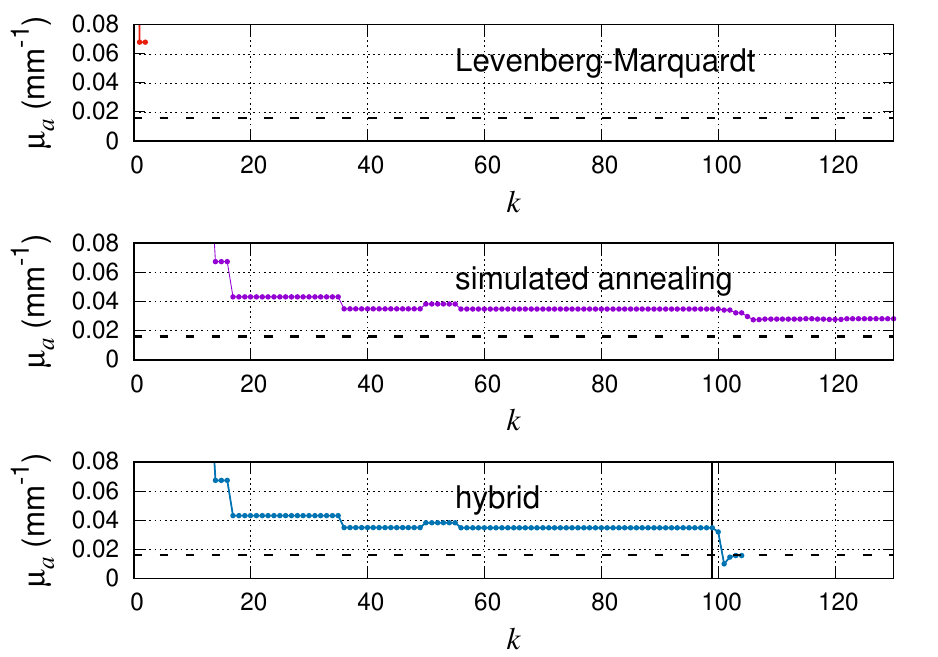}
\caption{\label{trs:fig3}
(Left) Case 2: Reconstructed $\mu_a$ by (top) Algorithm 1, (middle) Algorithm 2, and (bottom) Algorithm 3. The dashed lines show $\mu_a$ in Eq.~(\ref{correctanswer}). (Right) Same as the left panel but the vertical axes are from $0$ to $0.08$.
}
\end{figure}

\begin{figure}[ht]
\centering
\includegraphics[height=4.5cm]{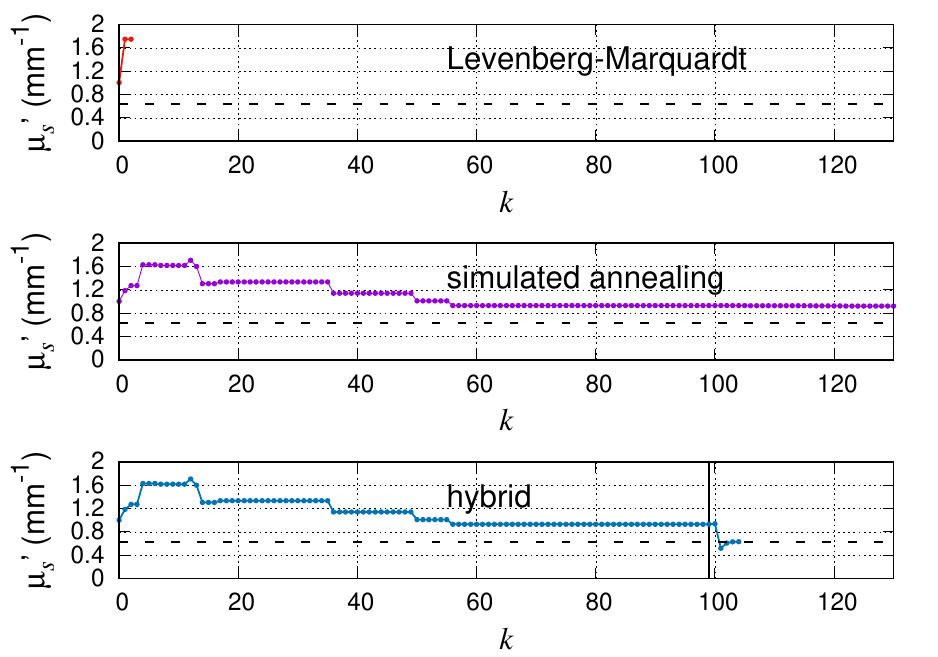}\hspace{-4mm}
\includegraphics[height=4.5cm]{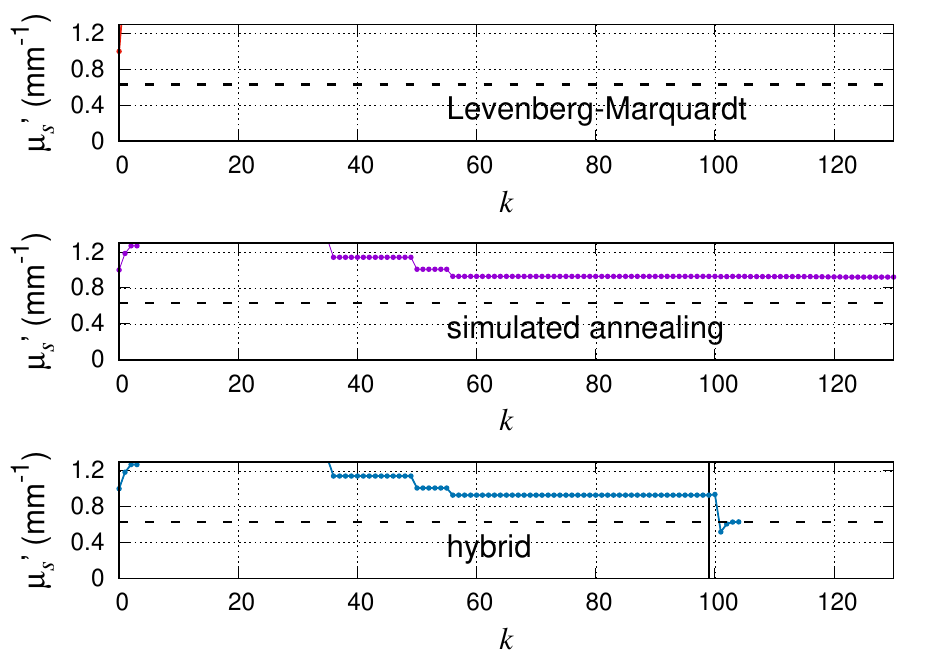}
\caption{\label{trs:fig4}
(Left) Case 2: Reconstructed $\mu_s'$ by (top) Algorithm 1, (middle) Algorithm 2, and (bottom) Algorithm 3. The dashed lines show $\mu_s'$ in Eq.~(\ref{correctanswer}). (Right) Same as the left panel but the vertical axes are from $0$ to $1.3$.
}
\end{figure}

\noindent
In this Case 2, Algorithm 1 (LM) fails and returns $\mu_a=0.068\,{\rm mm}^{-1}$ and $\mu_s'=1.75\,{\rm mm}^{-1}$ whereas Algorithm 3 (hybrid) still gives the correct values. The reconstructed values of $\mu_a$ and $\mu_s'$ at each iteration are shown in Figs.~\ref{trs:fig3} and \ref{trs:fig4}, respectively. In Fig.~\ref{trs:fig3}, the vertical axes of the left three panels are from $0$ to $0.6$ whereas the vertical axes of the right three panels are from $0$ to $0.08$. In Fig.~\ref{trs:fig4}, the vertical axes of the left three panels are from $0$ to $2$ while the vertical axes of the right three panels are from $0$ to $1.3$. In the top panels of Fig.~\ref{trs:fig3}, we see that the reconstruction of $\mu_a$ is unsuccessful by Algorithm 1 (LM). Algorithm 2 (SA) approaches $\mu_a$ in Eq.~(\ref{correctanswer}) but still deviates from that value in the middle panels of Fig.~\ref{trs:fig3}. As shown in the bottom panels of Fig.~\ref{trs:fig3}, Algorithm 3 (hybrid) converges in a few iterations after switching to Algorithm 1 (LM). In the top panels of Fig.~\ref{trs:fig4}, Algorithm 1 (LM) fails to reconstruct $\mu_s'$. In the middle panels of Fig.~\ref{trs:fig4}, reconstructed values by Algorithm 2 (SA) come close to $\mu_s'$ in Eq.~(\ref{correctanswer}) but suffer from slow convergence. In the bottom panels of Fig.~\ref{trs:fig4}, we see that Algorithm 3 (hybrid) arrives at $\mu_s'$ in Eq.~(\ref{correctanswer}) without any problem.

The initial guesses for Case 1 are reasonably close to the values found in Eq.~(\ref{correctanswer}). However, we started with initial guesses which are quite different from the above-mentioned values in Case 2. It is not surprising that Algorithm 1 (LM) does not work for Case 2, but Algorithm 3 (hybrid) can arrive at the correct values. The numerical calculations were performed on Matlab (i7-8700 CPU, 16 GB memory). In the hybrid method, the Metropolis-Hastings MCMC does not reach the steady state at about $100$ steps but can be switched to the Levenberg-Marquardt method. For Figs.~\ref{trs:fig3} and \ref{trs:fig4}, the computation time was $5\,{\rm sec}$. The simulated annealing method returns the correct value after about $10000$ steps, but it takes $8\,{\rm min}$. The computation for Algorithm 1 (LM) stopped in $0.5\,{\rm sec}$.

Below we summarize the reconstructed values on Table \ref{trs:tabla1}. Although Algorithm 2 (SA) and Algorithm 3 (hybrid) return the same results after a long time, the hybrid scheme is about ten times faster, and moreover there is no need of choosing the low temperature for the latter algorithm. For Algorithm 3, it is not necessary to wait until the burn-in time but it is enough if the initial rapid change ceases. In \S\ref{toy}, we illustrate that Algorithm 3 works once the Monte Carlo simulation escapes from a local minimum and the algorithm does not require that the calculation reaches the steady state.

\begin{table}[ht]
\centering
\begin{tabular}{lccccc}\hline
& Case 1 ($\mu_a$, $\mu_s'$) & Case 2 ($\mu_a$, $\mu_s'$) \\ \hline
%true                 & ($0.021$, $0.85$) & ($0.021$, $0.85$) \\
initial values       & ($0.01$, $1.0$) & ($0.5$, $1.0$) \\
Algorithm 1 (LM)     & ($0.016$, $0.63$) & ($0.068$, $1.75$) \\
Algorithm 2 (SA)     & ($0.032$, $1.02$)  & ($0.028$, $0.92$) \\
Algorithm 3 (hybrid) & ($0.016$, $0.63$) & ($0.016$, $0.63$)
\end{tabular}
\caption{\label{trs:tabla1}
Reconstructed values of $\mu_a,\mu_s'$ are shown for Case 1 and Case 2. The units of $\mu_a,\mu_s'$ are both ${\rm mm}^{-1}$. For Algorithm 2, values at $120$th Monte Carlo step are shown.
}
\end{table}

\subsection{Determination of absorption inhomogeneity}
\label{toy}

We perform a numerical experiment of diffuse optical tomography using the toy model.
Let us consider when the iterative scheme fails. We see $F'(0)=\left.\frac{\pp}{\pp a} u(\boldsymbol{\rho}^j,t^k;\boldsymbol{\rho}^i;a)\right|_{a=0}=0$. For sufficiently small $\epsilon>0$, we have $F'(\epsilon)>0$, $F'(-\epsilon)<0$, $|F'(\pm\epsilon)|\simeq0$, and 
$U-F(\pm\epsilon)<0$. Therefore if we start the iteration from $a_0=\epsilon$, we obtain

\be
|a_1|<\epsilon,\quad |a_2|<|a_1|,\quad \dots.
\ee

\noindent
Thus the sequence $a_k$ approaches $0$ and can never arrive at $\bar{a}$ ($=1.5$).

Let us consider how the difference $U-F(a)$ depends on $a$. We introduce

\be
h(t)=\frac{1}{t}\exp\left(-\frac{y_0^2}{4Dct}\right)
\left[1-\frac{\sqrt{\pi Dct}}{\ell}
e^{\left(\frac{y_0}{2\sqrt{Dct}}+\frac{\sqrt{Dct}}{\ell}\right)^2}
\mathrm{erfc}\left(\frac{y_0}{2\sqrt{Dct}}+\frac{\sqrt{Dct}}{\ell}\right)
\right].
\ee

\noindent
The following form is obtained using Eq.~(\ref{fwd:formula}). By neglecting noise, we have

\be
\begin{aligned}
&
u(\boldsymbol{\rho}^j,t^k;\boldsymbol{\rho}^i;\bar{a})-
u(\boldsymbol{\rho}^j,t^k;\boldsymbol{\rho}^i;a)
\\
&=
\frac{\eta e^{-\mu_{a0}ct^k}}{(2\pi D)^2}\int_0^{t^k}h(t^k-s)h(s)
\\
&\times
\left[\int_{-\infty}^{\infty}d_{\bar{a}}(a,x')\left(1-\tanh{{x'}^2}\right)
e^{-\frac{(x_d^j-x')^2}{4Dc(t^k-s)}}e^{-\frac{(x'-x_s^i)^2}{4Dcs}}
\,dx'\right]\,ds,
\end{aligned}
\ee

\noindent
where $d_{\bar{a}}(a,x')=\xi(\bar{a},x')-\xi(a,x')$ with

\be
\xi(a,x')=a^2\left[a+3\left(1+\frac{\tanh{{x'}^2}}{10}\right)\right].
\ee

\begin{figure}[ht]
\centering
\includegraphics[height=6cm]{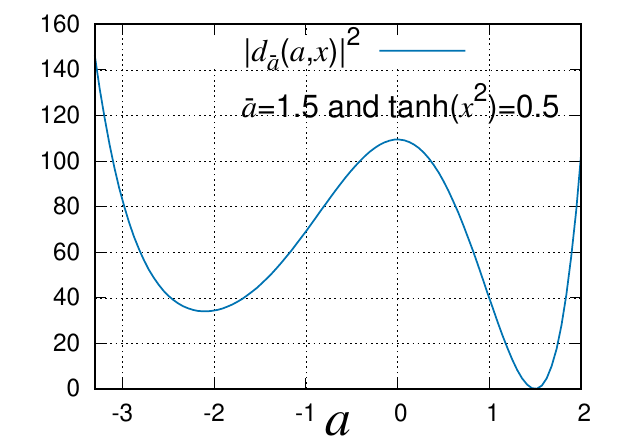}
\caption{\label{ot:figd}
The function $|d_{\bar{a}}(a,x)|^2$ ($\bar{a}=1.5$) is plotted when $\tanh(x^2)=0.5$.
}
\end{figure}

\noindent
For a given $x'$, the function $|d_{\bar{a}}(a;x')|^2$ has one global minimum at $a=\bar{a}$, one local minimum at $a=-2\left(1+\frac{\tanh{{x'}^2}}{10}\right)$, and one local maximum at $a=0$. Therefore, the above expression implies that Algorithm 1 (LM) fails when the initial value $a_0$ is negative and the correct value $\bar{a}=1.5$ is reconstructed only for a positive initial guess. Indeed, the computation ends up with $-2.05$ if we start the iteration from $a_0=-0.1$ (see below) or $-0.01$, and the value $1.68$ is obtained when $a_0=0.01$. Figure \ref{ot:figd} shows $|d_{\bar{a}}(a,x)|^2$ for $\bar{a}=1.5$ and $\tanh(x^2)=0.5$.

\begin{figure}[ht]
\centering
\includegraphics[height=6cm]{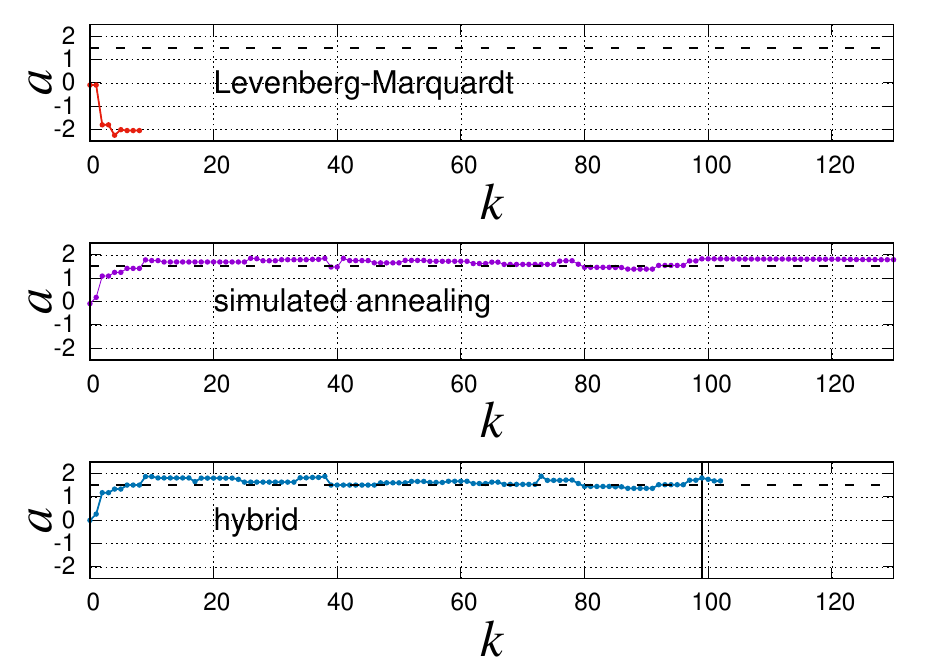}
\caption{\label{ot:fig}
Reconstructed $\bar{a}$ ($=1.5$) by (top) Algorithm 1 (LM), (middle) Algorithm 2 (SA), and (bottom) Algorithm 3 (hybrid) for the initial value $a_0=-0.1$.
}
\end{figure}

In Fig.~\ref{ot:fig}, we plot reconstructed values of $a$ against iteration numbers. The initial value is set to $a_0=-0.1$. The reconstruction by Algorithm 1 (LM) is shown in the top panel of Fig.~\ref{ot:fig}. As we can predict from Fig.~\ref{ot:figd}, Algorithm 1 (LM) converges to the local minimum and can never arrive at the global minimum. Monte Carlo simulation can jump over the local maximum and approach the global minimum, but keeps fluctuating as shown in the middle panel of Fig.~\ref{ot:fig}. We initially set $(\sigma_e,\varepsilon)=(10^{-6},0.5)$ for Algorithms 2 and 3. In Algorithm 2, we set $(\sigma_e,\varepsilon)=(10^{-7},0.005)$ after the {\em temperature} decreases at the $k_b$th Monte Carlo step. In the bottom panel of Fig.~\ref{ot:fig}, Algorithm 3 (hybrid) successfully arrives at the global minimum. Using Matlab, the computation time was $17\,{\rm min}$ while we needed $3\,{\rm hr}$ for Algorithm 2 (SA) ($1000$ steps).

After the initial time with large jumps, it is found that we can set $k_b=99$ in our simulation of the MCMC-iterative hybrid method. The correct value of $\bar{a}$ is reconstructed by Algorithm 3 (hybrid) even when the simulation experiences a local minimum. Starting from $a_{k_b}=1.8257$, the calculation stops at $a^*=1.6841$ ($a^*=a_{k_b+3}$). We note that the reconstructed $a^*$ is not exactly $1.5$ due to noise.

When the initial guess $a_0=-0.1$ lies in the valley of the local minimum (see Fig.~\ref{ot:figd}), the reconstructed $a$ by Algorithm 3 (hybrid) moves to the valley of the global minimum with the help of Monte Carlo simulation as shown in Fig.~\ref{ot:fig}, while the reconstructed $a$ by Algorithm 1 (LM) falls to the local minimum as the nature of Newton-type iterative methods. For Algorithm 3 (hybrid), we obtained $a^*=1.6841$. There is a possibility that negative reconstructed values are obtained by Algorithm 2 (SA) and Algorithm 3 (hybrid) for the first a few iterations if different pseudo-random numbers are used. These negative reconstructed values, however, will turn positive and the behavior of reconstructed $a$ by Algorithm 2 (SA) and Algorithm 3 (hybrid) is always more or less similar to the middle and bottom panels of Fig.~\ref{ot:fig}. For the initial guesses of $a_0=-0.01$ and $a_0=0.01$, Algorithm 3 (hybrid) works without any problem and Algorithm 2 (SA) also returns a reasonable result after a sufficiently large number of iterations (results not shown).

\section{Discussion}
\label{concl}

In this paper, we have proposed a hybrid numerical scheme which uses Markov chain Monte Carlo in the first step and then uses an iterative method in the second step. We switch from MH-MCMC to LM by observing proposed parameter values. For the typical jump length of parameters in MH-MCMC, $\varepsilon=0.1$ was used in Sec.~\ref{trs} and $\varepsilon=0.5$ was chosen in Sec.~\ref{toy}. Although these values were set so that the MH-MCMC calculation was efficiently performed, other choices of $\varepsilon$ are also possible. The proposed scheme is quite general and can be applied to different inverse problems in NIRS which are solved by iterative methods even when the forward problem has to be solved fully numerically with finite difference method or finite element method. It is an interesting future study how the hybrid scheme can be extended to diffuse optical tomography, which has a large number of unknowns.

More sophisticated algorithms than the Metropolis-Hastings Markov chain Monte Carlo used in this paper have been proposed in order to overcome slow convergence. The delayed rejection scheme reduces the net rejection rate \cite{Tierney94}. In the adaptive Metropolis algorithm, parameters in the proposal distribution are adjusted during Monte Carlo steps \cite{Haario-Saksman-Tamminen01}. The DRAM algorithm, which combines the above-mentioned two schemes, was also proposed \cite{Haario-Laine-Mira-Saksman06}. Two-level MCMC algorithms \cite{Christen-Fox05,Efendiev-Hou-Luo06} and a multi-level MCMC \cite{Langmore-Davis-Bal13} have been investigated to improve the MCMC algorithm. Such Monte Carlo schemes might improve the first step of our hybrid method by finding the valley of the global minimum more easily.

Related to the Metropolis-Hasitings Markov chain Monte Carlo algorithm, quantum annealing \cite{Kadowaki-Nishimori98} has been developed in addition to simulated annealing. Aiming at escaping from local minima, brute-force search and genetic algorithm \cite{Holland75} are also well-known optimization algorithms. The introduction of these methods in NIRS may be found useful in the future.

For the clinical use of NIRS, it has been recognized from early days that finding absolute values of the absorption and scattering coefficients is important \cite{Hoshi11}. In Ref.~\cite{Hebden-etal04}, it is emphasized that the obtained absolute values highly depend on the starting values of the initial estimate for the study of the infant brain with their measurement system. By performing Markov chain Monte Carlo before using standard iterative schemes, we may automatically acquire good initial values. Such clinical application is a natural next step of our research.

\section*{Acknowledgment}
%\ack 
The first author (Y.J.) is supported by National Natural Science Foundation of China (No. 11971121). Y.H. and M.M. acknowledge support from Grant-in-Aid for Scientific Research (17H02081) of the Japan Society for the Promotion of Science. M.M. also acknowledges support from Grant-in-Aid for Scientific Research (17K05572) of the Japan Society for the Promotion of Science and from the JSPS A3 foresight program: Modeling and Computation of Applied Inverse Problems. G.N. was supported by Grant-in-Aid for Scientific Research (15K21766 and 15H05740) of the Japan Society for the Promotion of Science. 
The authors appreciate Goro Nishimura for fruitful discussion on optical tomography and the use of his computer facilities at Hokkaido University. We thank Tatsunori Emi, Takeshi Iwasaki, and Tomoya Sugiyama for helping related experiments which stimulated the present work. 

%\newpage
%\setcounter{section}{1}
%\appendix

%\section*{References}

\end{document}